\documentclass{pasj01}
\draft

\SetRunningHead{Ohta et al.}{Extreme Asymmetry in the Disk of V1247~Ori}
\Received{}
\Accepted{}
\Published{}

\begin{document}

\title{Extreme Asymmetry in the Polarized Disk of V1247~Ori\thanks{Based on data collected at Subaru Telescope, which is operated by the National Astronomical Observatory of Japan.}}

\author{%
   Yurina \textsc{Ohta}\altaffilmark{1}, 
   Misato \textsc{Fukagawa} \altaffilmark{1,2,3}, 
   Michael L. \textsc{Sitko}\altaffilmark{4,5,6}, 
   Takayuki \textsc{Muto}\altaffilmark{7}, 
   Stefan \textsc{Kraus}\altaffilmark{8}, 
   Carol A. \textsc{Grady}\altaffilmark{9,10}, 
   John P. \textsc{Wisniewski}\altaffilmark{11}, 
   Jeremy R. \textsc{Swearingen}\altaffilmark{4}, 
   Hiroshi \textsc{Shibai}\altaffilmark{1}, 
   Takahiro \textsc{Sumi}\altaffilmark{1}, 
   Jun \textsc{Hashimoto}\altaffilmark{2}, 
   Tomoyuki \textsc{Kudo}\altaffilmark{12}, 
   Nobuhiko \textsc{Kusakabe}\altaffilmark{2}, 
   Munetake \textsc{Momose}\altaffilmark{13}, 
   Yoshiko \textsc{Okamoto}\altaffilmark{13}, 
   Takayuki \textsc{Kotani}\altaffilmark{2}, 
   Michihiro \textsc{Takami}\altaffilmark{14}, 
   Thayne \textsc{Currie}\altaffilmark{12}, 
   Christian \textsc{Thalmann}\altaffilmark{15},  
   Markus \textsc{Janson}\altaffilmark{16}, 
   Eiji \textsc{Akiyama}\altaffilmark{2}, 
   Katherine B. \textsc{Follette}\altaffilmark{17}, 
   Satoshi \textsc{Mayama}\altaffilmark{18}, 
   Lyu \textsc{Abe}\altaffilmark{19}, 
   Wolfgang \textsc{Brandner}\altaffilmark{20}, 
   Timothy D. \textsc{Brandt}\altaffilmark{21}, 
   Joseph C. \textsc{Carson}\altaffilmark{22}, 
   Sebastian E. \textsc{Egner}\altaffilmark{12}, 
   Markus \textsc{Feldt}\altaffilmark{20}, 
   Miwa \textsc{Goto}\altaffilmark{23}, 
   Olivier \textsc{Guyon}\altaffilmark{12},
   Yutaka \textsc{Hayano}\altaffilmark{12}, 
   Masahiko \textsc{Hayashi}\altaffilmark{2}, 
   Saeko S. \textsc{Hayashi}\altaffilmark{12}, 
   Thomas \textsc{Henning}\altaffilmark{20}, 
   Klaus W. \textsc{Hodapp}\altaffilmark{24}, 
   Miki \textsc{Ishii}\altaffilmark{12}, 
   Masanori \textsc{Iye}\altaffilmark{2}, 
   Ryo \textsc{Kandori}\altaffilmark{2}, 
   Gillian R. \textsc{Knapp}\altaffilmark{21}, 
   Masayuki \textsc{Kuzuhara}\altaffilmark{25}, 
   Jungmi \textsc{Kwon}\altaffilmark{26}, 
   Taro \textsc{Matsuo}\altaffilmark{1}, 
   Michael W. \textsc{McElwain}\altaffilmark{10}, 
   Shoken \textsc{Miyama}\altaffilmark{27}, 
   Jun-Ichi \textsc{Morino}\altaffilmark{2}, 
   Amaya \textsc{Moro-Mart\'in}\altaffilmark{28,29}, 
   Tetsuo \textsc{Nishimura}\altaffilmark{12}, 
   Tae-Soo \textsc{Pyo}\altaffilmark{12}, 
   Eugene \textsc{Serabyn}\altaffilmark{30}, 
   Takuya \textsc{Suenaga}\altaffilmark{31}, 
   Hiroshi \textsc{Suto}\altaffilmark{12}, 
   Ryuji \textsc{{Suzuki}\altaffilmark{12}, 
   Yasuhiro H. \textsc{Takahashi}\altaffilmark{26}, 
   Hideki \textsc{Takami}\altaffilmark{12}, 
   Naruhisa \textsc{Takato}\altaffilmark{12}, 
   Hiroshi \textsc{Terada}\altaffilmark{2}, 
   Daigo \textsc{Tomono}\altaffilmark{12}, 
   Edwin L. \textsc{Turner}\altaffilmark{21,32}, 
   Tomonori \textsc{Usuda}\altaffilmark{2}, 
   Makoto \textsc{Watanabe}\altaffilmark{33}, 
   Toru \textsc{Yamada}\altaffilmark{34},}
   \and Motohide \textsc{Tamura}\altaffilmark{2,26}
    }
\altaffiltext{1}{Graduate School of Science, Osaka University, 1-1 Machikaneyama, Toyonaka, Osaka 560-0043, Japan}\email{ohta@iral.ess.sci.osaka-u.ac.jp}
\altaffiltext{2}{National Astronomical Observatory of Japan, 2-21-1, Osawa, Mitaka, Tokyo 181-8588, Japan}
\altaffiltext{3}{Division of Particle and Astrophysical Science, Graduate School of Science, Nagoya University, Furo-cho, Chikusa-ku, Nagoya, Aichi 464-8602, Japan}
\altaffiltext{4}{Department of Physics, University of Cincinnati, Cincinnati, OH 45221, USA}
\altaffiltext{5}{Space Science Institute, 475 Walnut Street, Suite 205, Boulder, CO 80301, USA}
\altaffiltext{6}{Visiting Astronomer, NASA Infrared Telescope Facility, operated by the University of Hawaii under contract with the National Aeronautics and Space Administration.}
\altaffiltext{7}{Division of Liberal Arts, Kogakuin University, 1-24-2, Nishi-Shinjuku, Shinijuku-ku, Tokyo 163-8677, Japan}
\altaffiltext{8}{School of Physics, University of Exeter, Stocker Road, Exeter EX4 4QL, UK}
\altaffiltext{9}{Eureka Scientific Inc., Oakland, CA 94602, USA}
\altaffiltext{10}{Exoplanets and Stellar Astrophysics Laboratory, Code 667, Goddard Space Flight Center, Greenbelt, MD 20771, USA}
\altaffiltext{11}{H. L. Dodge Department of Physics and Astronomy, University of Oklahoma, 440 W. Brooks St Norman, OK 73019, USA}
\altaffiltext{12}{Subaru Telescope, 650 North A’ohoku Place, Hilo, HI 96720, USA}
\altaffiltext{13}{College of Science, Ibaraki University, 2-1-1 Bunkyo, Mito, Ibaraki 310-8512 Japan}
\altaffiltext{14}{Institute of Astronomy and Astrophysics, Academia Sinica, P.O. Box 23-141, Taipei 10617, Taiwan, R.O.C}
\altaffiltext{15}{Astronomical Institute "Anton Pannekoek", University of Amsterdam, Postbus 94249, 1090 GE, Amsterdam, The Netherlands}
\altaffiltext{16}{Department of Astronomy, Stockholm University, AlbaNova University Center, 10691 Stockholm, Sweden}
\altaffiltext{17}{Kavli Institute of Particle Astrophysics and Cosmology, Stanford University, 452 Lomita Mall, Stanford, CA 94305, USA}
\altaffiltext{18}{The Center for the Promotion of Integrated Sciences, The Graduate University for Advanced
Studies (SOKENDAI), Shonan International Village, Hayama-cho, Miura-gun, Kanagawa 240-0193}
\altaffiltext{19}{Laboratoire Lagrange (UMR 7293), Universit\'e de Nice-Sophia Antipolis, CNRS, Observatoire de la C\^ote d'Azur, 28 avenue Valrose, 06108 Nice Cedex 2, France}
\altaffiltext{20}{Max Planck Institute for Astronomy, K\"{o}nigstuhl 17, D-69117 Heidelberg, Germany}
\altaffiltext{21}{Department of Astrophysical Science, Princeton University, Peyton Hall, Ivy Lane, Princeton, NJ 08544, USA}
\altaffiltext{22}{Department of Physics and Astronomy, College of Charleston, 58 Coming St., Charleston, SC 29424, USA}
\altaffiltext{23}{Universit\"ats-Sternwarte M\"unchen, Ludwig-Maximilians-Universit\"at, Scheinerstr. 1, 81679 M\"unchen, Germany}
\altaffiltext{24}{Institute for Astronomy, University of Hawaii, 640 N. A'ohoku Place, Hilo, HI 96720, USA}
\altaffiltext{25}{Department of Earth and Planetary Sciences, Tokyo Institute of Technology, 2-12-1 Ookayama, Meguro-ku, Tokyo 152-8551, Japan}
\altaffiltext{26}{The University of Tokyo, 7-3-1, Hongo, Bunkyo-ku, Tokyo 113-0033, Japan}
\altaffiltext{27}{Hiroshima University, 1-3-1, Kagamiyama, Higashi-Hiroshima, Hiroshima 739-8526, Japan}
\altaffiltext{28}{Space Telescope Science Institute, 3700 San Martin Drive, Baltimore, MD, 21218, USA}
\altaffiltext{29}{Center for Astrophysical Sciences, Johns Hopkins University, Baltimore MD 21218, USA}
\altaffiltext{30}{Jet Propulsion Laboratory, California Institute of Technology, Pasadena, CA, 91109, USA}
\altaffiltext{31}{The Graduate University for Advanced Studies (SOKENDAI), 2-21-1 Osawa, Mitaka, Tokyo 181-8588, Japan}
\altaffiltext{32}{Kavli Institute for Physics and Mathematics of the Universe, The University of Tokyo, 5-1-5, Kashiwanoha, Kashiwa, Chiba 277-8568, Japan}
\altaffiltext{33}{Department of Cosmosciences, Hokkaido University, Kita-ku, Sapporo, Hokkaido 060-0810, Japan}
\altaffiltext{34}{Astronomical Institute, Tohoku University, Aoba-ku, Sendai, Miyagi 980-8578, Japan}


\maketitle

\begin{abstract}
We present the first near-infrared scattered-light detection of the transitional disk around V1247~Ori, which was obtained using high-resolution polarimetric differential imaging observations with Subaru/HiCIAO. 
Our imaging in the $H$ band reveals the disk morphology at separations of $\sim0\farcs14$--$0\farcs86$ (54--330~au) from the central star.
The polarized intensity ($PI$) image shows a remarkable arc-like structure toward the southeast of the star, whereas the fainter northwest region does not exhibit any notable features. 
The shape of the arm is consistent with an arc of $0\farcs28 \pm 0\farcs09$ in radius (108 au from the star), although the possibility of a spiral arm with a small pitch angle cannot be excluded. 
V1247~Ori features an exceptionally large azimuthal contrast in scattered, polarized light; the radial peak of the southeastern arc is about three times brighter than the northwestern disk measured at the same distance from the star. 
Combined with the previous indication of an inhomogeneous density distribution in the gap at $\lesssim$46~au, the notable asymmetry in the outer disk suggests the presence of unseen companions and/or planet-forming processes ongoing in the arc.
\end{abstract}

\section{Introduction}
Protoplanetary disks are known to be the sites of planet formation and have been extensively studied in recent years.
One class of disks in particular has received much attention:  (pre-)transitional disks whose spectral energy distributions (SEDs) show a deficit in the near- to mid-infrared, suggesting a lack of grains at the corresponding temperature/radius in the disks \citep{str89,cal02}. 
In fact, the gaps and holes have been spatially resolved over a  wide range of wavelengths in millimeter and infrared (e.g., \cite{and11,has12,may12,qua13}). 
Even more characteristic are the spiral arms found in (pre-)transitional disks,  although the detections are limited for a small number of intermediate-mass ($\sim 2M_{\solar}$) stars (e.g., SAO~206462, \cite{gar13,mut12}; MWC~758, \cite{ben15,gra13}; HD~142527, \cite{cas12,ave14a,rod14}; HD~100546, \cite{boc13,cur14,ave14b}). 
Because planets may account for these unique structures, increasing the sample size of spatially-resolved observations is important to obtain an insight into planet-disk interaction as well as  planet formation in action. 

V1247~Ori (HD~290764) is a pre-main sequence star located at $d=385$$\pm$15~pc \citep{cab08,ter07} with an F0V spectral type \citep{vie03,kra13}, and its stellar mass and age have been estimated as $M_*=1.86$$\pm$0.02~$M_{\odot}$ and 7.4$\pm$0.4~Myr, respectively \citep{kra13}. 
Its SED exhibits a dip in the wavelength range of 3--15~$\mu$m similar to SAO~206462 \citep{bro09,sit12}, indicating that the star is surrounded by a (pre-)transitional disk \citep{cab10}. 
The disk is considered to be optically thick in the region beyond the expected radial gap, judging from the amount of the infrared excess (the excess luminosity is on the order of 0.1 of the stellar luminosity).
The first high-angular-resolution observations were obtained by \citet{kra13} who performed infrared interferometry at 1.5--13~$\mu$m with the VLTI and Keck long-baseline interferometers, as well as Gemini and Keck aperture masking to explore the inner hot region of the disk with sub-au angular resolution. 
The best estimates of the disk inclination ($31\fdg3 \pm 7\fdg5$) and position angle (P.A.) ($104\arcdeg \pm 15\arcdeg$) were obtained at mid-infrared wavelengths (8--12~$\mu$m) for the disk's thermal emission within 75 mas (29 au). 
Notably, they found the system consisting of a dust disk near the dust sublimation radius (0.18~au), an outer disk beyond 46~au, and a radial, partial gap between them filled with optically thin material.
The material in the gap region was found  to be inhomogeneously distributed, as indicated by strong non-zero phase signals measured with aperture masking interferometry. 

In this paper, we present the results of the polarimetric differential imaging (PDI) of V1247~Ori at high angular resolution ($0\farcs07=27$~au) in the near-infrared, obtained to uncover the detailed structure of the outer disk where the interferometric data mentioned above are insufficiently sensitive. 
PDI is a powerful technique to directly image circumstellar structures in light scattered by dust grains while suppressing the stellar light. 
In our observations, an arc-like structure is discovered in the region within $\sim$200~au from the central star, causing the strong azimuthal asymmetry to be found for the first time in this outer disk. 

\section{Observations}
V1247 Ori ($H=8.20$; \cite{cut03}) was targeted first on November 23, 2013. 
The sky was clear on that night. The observations were done as part of the Strategic Explorations of Exoplanets and Disks with Subaru (SEEDS; \cite{tam09}) using the HiCIAO camera \citep{tam06} and the adaptive optics AO188 \citep{hay04} on the Subaru telescope.
The PDI mode was used in the $H$ band (1.6~$\mu$m) in combination with angular differential imaging (ADI). 
The guiding star for the adaptive optics was V1247~Ori itself ($\sim$10.0 mag in $R$). 
A spatial resolution (FWHM) of $0\farcs07$ was achieved on average, as measured using unsaturated snap shots of V1247~Ori with the neural density (ND) filter (transmission of 9.740\%$\pm$0.022\%) taken just before the disk imaging. 
A Lyot-type coronagraph was used to block out the light from the central star with an occulting mask  $0\farcs3$ in diameter. 
The single PDI (sPDI) observing mode was employed whereby the incident light was divided into two beams by a Wollaston prism with linear polarization states perpendicular to each other. 
These two images were simultaneously obtained in one data frame and the field of view of each image was  about $10\arcsec \times 20\arcsec$ (1024 $\times$ 2048 pixels). 
The half-wave plate (HWP) was rotated to position angles of $0\arcdeg$, $45\arcdeg$, $22\fdg5$, and $67\fdg5$ in sequence, and one frame was obtained at each angle. 
This cycle, consisting of the four position angles of the HWP, was repeated 21 times; thus, 84 frames were obtained. 
The exposure time was 30 seconds per frame, and 42 minutes in total. 
The total field rotation for ADI was $31\fdg8$. 
Just after the imaging of V1247~Ori, HD~288196 ($H=5.95$; \cite{cut03}) was observed as a point spread function (PSF) reference star. 
HD~288196 produced a similar illumination of the AO wave front sensor 
through the ND filter as did V1247~Ori, and the data were obtained with the same AO settings.
The sPDI+ADI mode was used in the $H$ band with the same $0\farcs3$ mask. 
Three HWP cycles were obtained, producing a total of 12 data frames. 
The exposure time for each frame was 10 seconds. 
The photometric calibration was done using the data of the standard star HD~44612 observed on the same night in sPDI. 

The follow-up observations were performed two months later, on January 19 in 2014 through thin cirrus clouds. 
The same $H$ filter was used with HiCIAO+AO188, but at this epoch, the quadruple PDI (qPDI) mode was chosen for the imaging polarimetry without an occulting mask to probe a closer-in region than in the previous sPDI observations. 
In qPDI, the incident light was separated into 4 beams by double Wollaston prisms and the central star was less saturated compared to the sPDI mode. 
In this case, two sets of images with orthogonal polarization states were recorded in one data frame, and the field of view was approximately $6\arcsec \times 6\arcsec$ ($640 \times 640$~pixels). 
A FWHM of $0\farcs069$ 
was achieved for V1247~Ori, estimated using the unsaturated images through the ND filter taken before the disk imaging. 
However, the snap shots after the disk imaging indicated a FWHM 1.2 times larger than that. 
It was also noticeable that the halo of the PSFs became blurred with time during long exposures for the disk, and such frames were discarded in the process of data reduction as described in the next section. 
Allowing saturation around the stellar peak for a radius of 2 pixels, three exposures of 10 sec were co-added into one frame, giving 30 sec as the total exposure time per frame.
Eighteen HWP cycles were obtained and 72 frames were acquired, corresponding to a total integration time of 36 minutes. The ADI was combined as in the previous epoch and the field rotation was $29\fdg9$. 
No PSF reference was obtained. 
The photometric standard star, HD~129655, was observed on the same night in qPDI.

\section{Data Reduction}
The data reduction was performed using the IRAF (Image Reduction and Analysis Facility) software packages. 
The first noise to be corrected was the horizontal and vertical stripe patterns  in  raw images of HiCIAO \citep{suz10}. 
These stripes were eliminated in the same manner as used by \citet{yam13}.
The dark image was used to identify hot and warm pixels. 
Hot pixels were marked if the pixel values exceeded 150$\sigma_s$ where $\sigma_s$ was a standard deviation measured using all pixels in the dark frame after 5$\sigma_s$ clipping. 
The warm pixels were found next to the hot ones, showing significantly larger values than the surrounding normal pixels though not quite as high as the hot pixels (roughly about 1\% of the central hot pixel, or over 8$\sigma_s$). 
Those pixels were substituted by bi-linear interpolation. 
The flat-fielding was performed using dome-flats. 
One data frame was divided into two and four for sPDI and qPDI, respectively, and corrected for the distortion by comparing the HiCIAO data taken in the same observing run  and the image taken by the Hubble Space Telescope for the globular clusters M15 and M5. 
The pixel scale after the distortion correction was 9.5 mas pixel$^{-1}$. 
Further position matching among the multiple images was done by taking  cross-correlation of the halo of the central star. 

The unpolarized light was subtracted by differencing the two orthogonal polarization components. 
The Stokes parameters, $+Q, -Q, +U, -U$, were obtained as follows: $+Q = F_{\rm 0\degree} - F_{\rm 90\degree}$, $-Q = F_{\rm 90\degree} - F_{\rm 0\degree}$, $+U = F_{\rm 45\degree} - F_{\rm 135\degree}$, $-U = F_{\rm 135\degree} - F_{\rm 45\degree}$, where $F_{\theta}$ is the flux density for the light polarized at $\theta$ which is twice the position angle of the HWP. 
The double differential technique was then used to further reduce the residuals, where $Q = \{+Q - (-Q)\}/2$ and $U = \{+U - (-U)\}/2$. 
The smearing of the images was unavoidable when performing the double differential because of the field rotation over the 2 HWP angles, but the amount of the rotation was very small: $0\fdg6$--$0\fdg7$ for sPDI, and $0\fdg5$--$0\fdg8$ for qPDI. 
After correcting  for instrumental polarization caused by HiCIAO at the Nasmyth focus (e.g.,  \cite{joo08}), the images of the $Q$ and $U$ Stokes parameters were de-rotated to match the field orientation. 
The de-rotated images were averaged over multiple HWP cycles after two and five HWP cycles were discarded for sPDI and qPDI, respectively, because of the larger sizes of the PSFs compared with other cycles. 
The polarized intensity ($PI$) was calculated as $PI=\sqrt{Q^2 + U^2}$. 
The uncertainty ($\sigma$) was estimated by propagation from the standard deviations of $Q$ and $U$ over the HWP cycles. 
We also checked the consistency between the two sets of Stokes parameters obtained using two sets of orthogonal beams for the qPDI. 
There was a systematic difference because of the mismatch of the PSF shape, possibly caused by an uncertainty of distortion correction.
The resulting $PI$ differed by 10--18\% depending upon the location in the disk, but this does not significantly affect the discussion in later sections. 
The stellar position was estimated as the brightness centroid of the PSF halo. 
The centroid was calculated in the region within $1\farcs4$ from the center of the occulting mask for sPDI.
It was estimated in an aperture of $r=0\farcs07$ (1 FWHM, 7 pixels) after artificially masking the saturated area at $r\leq2$ pixels in qPDI. 
The uncertainty in the stellar position was 0.3 pixels (3~mas).

\begin{figure}
 \begin{center}
  \includegraphics[width=16cm]{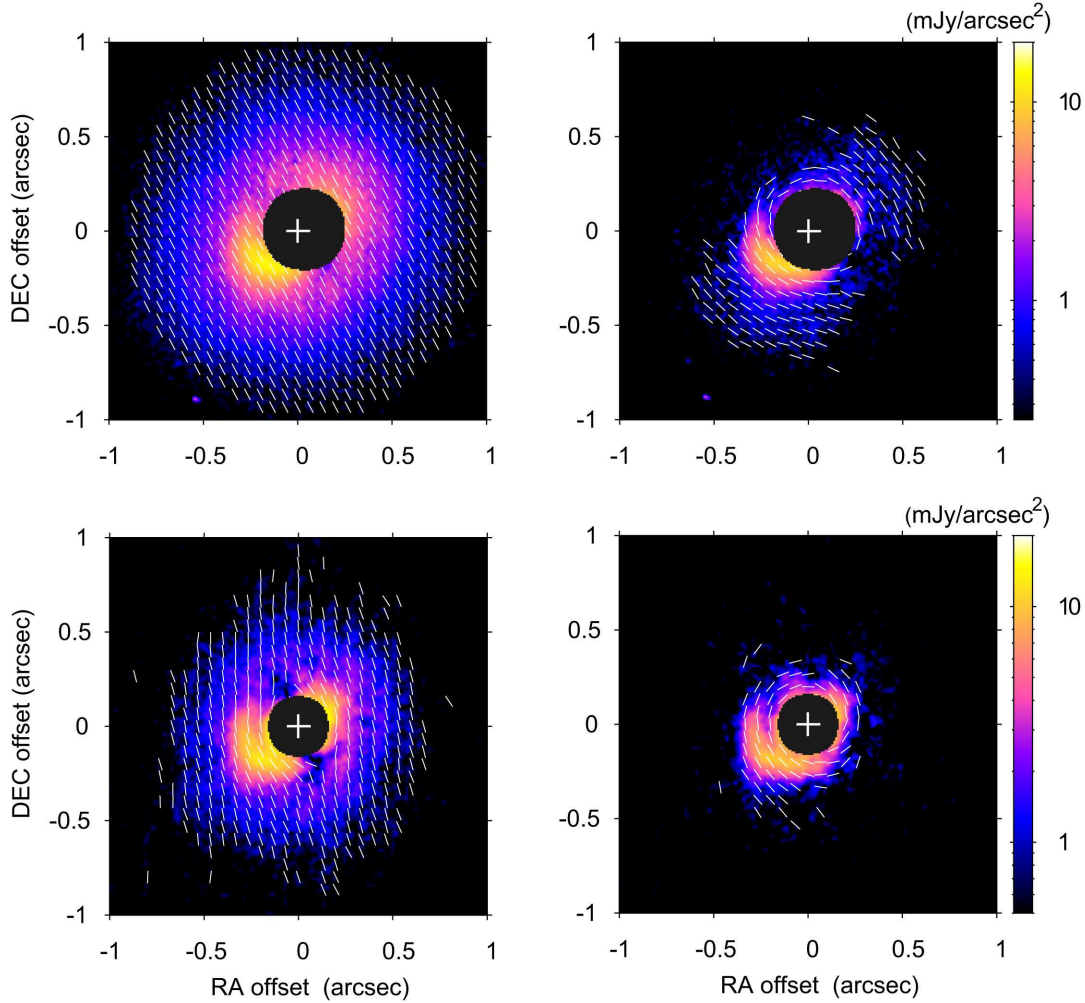}
 \end{center}
\caption{Polarized intensity images for V1247~Ori, overlaid with polarization vectors. The stellar position is centered at (0, 0) and marked with a cross. The $left$ panels present the $PI$ and polarization angle before the subtraction of the polarized PSF whereas the $right$ ones show those images after subtraction. The $upper$ and $lower$ images were obtained in sPDI and qPDI, respectively. The polarization angle was calculated using the $Q$ and $U$ images with a binning of $7\times7$ pixels (1 FWHM). The vector lengths are constant (arbitrary) and do not represent the polarization degree. The vectors are plotted in the region where the SNR of $PI$ is greater than or equal to 3. The central {\it black circle} depicts the region that was obscured by the coronagraphic mask or that is dominated by residuals of the PSF, with a radius of $0\farcs2$ for sPDI or $0\farcs14$ for qPDI. North is up and East is to the left.}\label{fig:polzhalo}
\end{figure}

A spatially extended, halo-like $PI$ distribution was seen in the reduced images. 
The polarization position angle was $\sim$29$\arcdeg$ in sPDI and $\sim$10$\arcdeg$ in qPDI everywhere, regardless of the direction from the central star (Figure~\ref{fig:polzhalo}). 
This suggests that the PSF itself was polarized (e.g., \cite{fol15}). 
We have attempted to separate this unresolved (PSF) polarized component at a constant direction to explore whether another polarized component remains after subtracting it. 
Assuming that the contribution from the disk to the Stokes $I$ was negligible compared to that from the central star, an artificial, polarized PSF was constructed for $Q$ and $U$ as  $Q_{\rm psf} = P_{\rm psf}I\cos{\theta}_{\rm psf}$, $U_{\rm psf} = P_{\rm psf}I\sin{\theta}_{\rm psf}$. 
$Q_{\rm psf}$ and $U_{\rm psf}$ were then subtracted from the observed $Q$ and $U$ to calculate the $PI$. 
This process was iterated by varying $P_{\rm psf}$ and ${\theta}_{\rm psf}$ in steps of $\delta P_{\rm psf}=0.05\%$ and $\delta{\theta}_{\rm psf}=0\fdg5$ until the aligned polarization in the halo-like $PI$ was minimized in the region of P.A.=0$\arcdeg$(180$\arcdeg$)--40$\arcdeg$(220$\arcdeg$). 
The inferred polarization of the PSF was $P_{\rm psf} = 0.85\pm0.05$~\% and  $\theta_{\rm psf} = 29\arcdeg \pm 2\arcdeg$ for sPDI, and $P_{\rm psf} = 0.80\pm0.05$~\% and  $\theta_{\rm psf} = 12\arcdeg \pm 2\arcdeg$ for qPDI. 
The distribution of the polarization angle is shown in the right panels of Figure~\ref{fig:polzhalo}. 

The polarized PSF can be accounted for by the scattering in an unresolved, inner portion of the disk. 
This interpretation is consistent with the fact that the PSF was not polarized for the reference star obtained in sPDI. 
The $PI$ was not significantly detected and the random orientation of the polarization was found at the noise level ($<$3$\sigma$) for the reference star. 
In addition, the polarized PSF has been common in HiCIAO observations for inclined disks, and the direction of polarization is always similar to that of the disk minor axis (e.g., \cite{fol15}). 
The polarization degree is expected to be higher toward the region running along the disk major axis, where the scattering angle is closer to $90\arcdeg$ than in the region of the minor axis, and hence the net polarization in a resolution element centered on a star should be dominated by the scattered light polarized perpendicular to the direction of the major axis. 
V1247~Ori is not an exception, because the observed polarization direction of the halo-like component is consistent with the P.A. of the disk minor axis within the uncertainty ($14\arcdeg\pm15\arcdeg$, \cite{kra13}). 
It should be noted, however, that the polarization degree of the PSF is small, and the angle cannot be used to accurately estimate the P.A. of the disk minor or major axis because it could readily be affected by instrumental effects such as PSF shape and a coronagraphic mask (metal film). 

The aperture photometry for V1247 Ori was performed using the unsaturated short exposures in our observations. 
The aperture radius was 100 pixels for sPDI and 50 pixels for qPDI. 
The resulting magnitudes were $H=8.15 \pm 0.04$ and $8.26 \pm 0.04$ at the epochs of sPDI and qPDI, respectively. 
Considering that the qPDI photometry may slightly underestimate the magnitudes because of the lower signal-to-noise ratio (SNR) for the PSF halo, these values are not inconsistent and also agree with the results of 2MASS photometry of $H=8.20\pm0.05$, obtained on October 30, 1998 \citep{cut03}. 

In the process of deriving the $PI$, no assumption was made on the polarization direction for the circumstellar material that could be spatially resolved. 
However, $PI$ is always positive and biased by positive noise.
Thus, the technique to extract the azimuthal component has often been used recently, assuming that the disk scattered light is polarized perpendicular to the direction toward the central star (an illuminating source). 
The azimuthal component was computed as $Q_{\phi} = Q \times cos{(2\phi)} + U \times sin{(2\phi)}$, where $\phi$ is the position angle. 
It was confirmed that $PI$ and $Q_{\phi}$ were consistent with each other within the uncertainty $\sigma$ both for the sPDI and qPDI observing modes.

\section{Results}
\subsection{Disk Morphology}
\begin{figure}
 \begin{center}
  \includegraphics[width=16cm]{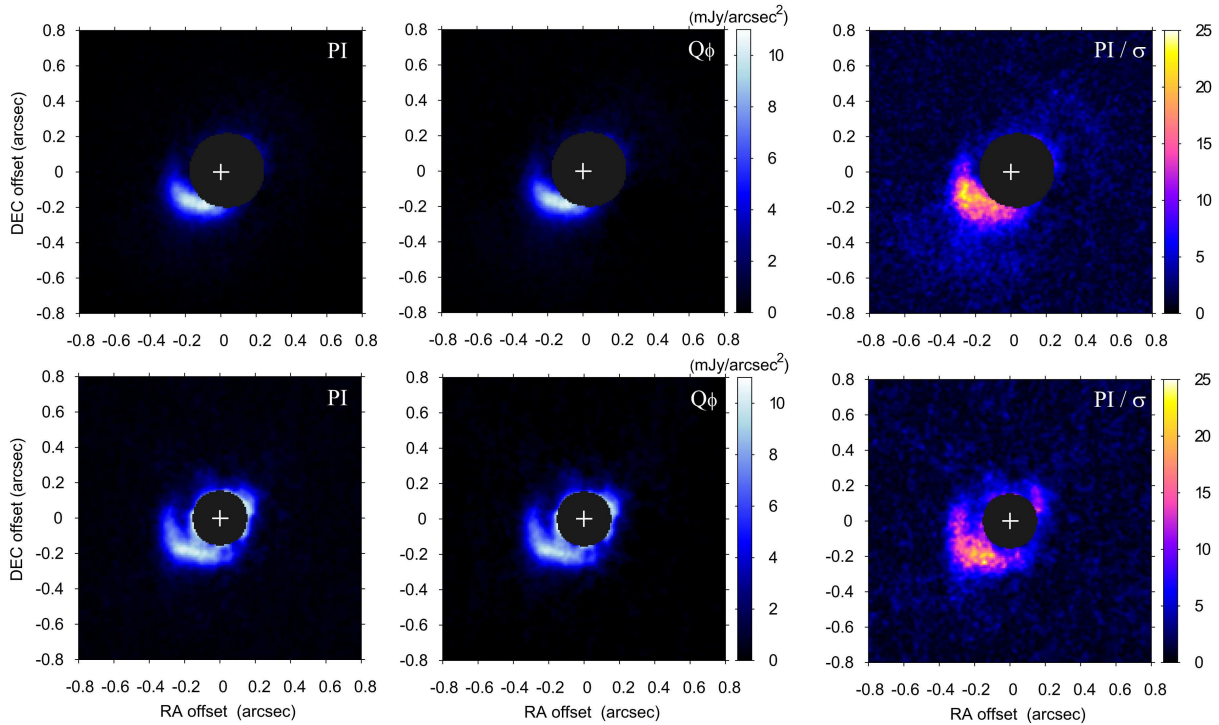}
 \end{center}
\caption{Polarized intensity images of V1247~Ori in the $H$ band, measured in the sPDI ($top$) and qPDI ($bottom$) modes. The stellar position is denoted with a cross. The $PI$ ($left$), $Q_{\phi}$ ($middle$), and $PI$/$\sigma$ ($right$) maps are shown. The display range is the same for $PI$ and $Q_{\phi}$ images.}\label{fig:pi}
\end{figure}

Polarized light was detected around V1247 Ori, and the polarization direction indicated that the circumstellar disk was spatially resolved in scattered light in our observations (Figure~\ref{fig:polzhalo}, \ref{fig:pi}). 
The $PI$ was detected over the $3\sigma$ level out to a radius ($r$) of $0\farcs66$ (254~au) and $0\farcs86$ (330~au) in the southeastern  (SE) and northwestern (NW) directions, respectively, in the sPDI with a higher SNR than the qPDI. 
Here we define the position angles of SE and NW as $119\arcdeg$ and $299\arcdeg$, respectively, given the previous estimate of the major-axis P.A. as $104\arcdeg \pm 15\arcdeg$ \citep{kra13} and the recent Submillimeter Array observations that favor the larger end of this range for the outer disk (Kraus et al. in private communication). 

The inner working angle (IWA) achieved is $0\farcs16$ (62~au) in SE and $0\farcs24$ (92~au) in the NW for sPDI because the center of the coronagraphic mask was displaced from the stellar position by $0\farcs04$ toward the SE and the area within $0\farcs2$ of the mask center suffered from obscuration by the occulting mask. 
In the qPDI image taken without a mask, the IWA was estimated to be $0\farcs14$ (54~au), because the residuals of the stellar light were significant and the SNR was below than 10 within that radius. 
Scattered light can be seen in the qPDI image at $r \lesssim 0\farcs2$ ($\sim$80~au) just outside of the IWA. 
This must be confirmed by future imaging with a better IWA, but considering the  polarization angle in this radial range, which appears point-symmetric as expected by light scattering, this might be part of the outer disk at $r > 46$~au, as suggested by \citet{kra13}. 

Most remarkable is the bright, arc-like structure toward the SE found for the first time in this disk. 
The integrated $PI$ in $r=0\farcs24$--$0\farcs9$ is 0.53$\pm$0.10 mJy, which is 0.096\% 
of the total intensity (Stokes $I$) of V1247~Ori in the $H$ band. 
If measured in the SE region including the arc ($0\farcs24 \leq r \leq 0\farcs90$, $29\arcdeg \leq$ P.A. $\leq 209\arcdeg$), the total $PI$ is 0.43$\pm$0.07~mJy (81\% of the whole disk), whereas it is 0.097$\pm$0.025~mJy (18\%) in the NW ($0\farcs24 \leq r \leq 0\farcs90$, $-151\arcdeg \leq$ P.A. $\leq  29\arcdeg$). 
The consistent results were obtained in the same measurements but using the $Q_{\phi}$ image.

Note that the elliptical isophote fitting suggests $137\fdg2 \pm 3\fdg2$ as the P.A. of the disk major axis when estimated in the $0\farcs4 < r < 0\farcs6$ region of the sPDI. 
It is also possible that the P.A. differs between the inner disk probed by interferometry \citep{kra13} and the outer disk discussed in this work. 
However, we assume the P.A. measured in \citet{kra13} in this paper, because the SE arc may bias our estimate of the P.A.

\subsection{Radial Surface Brightness Distribution}
\begin{figure}
 \begin{center}
  \includegraphics[width=10cm]{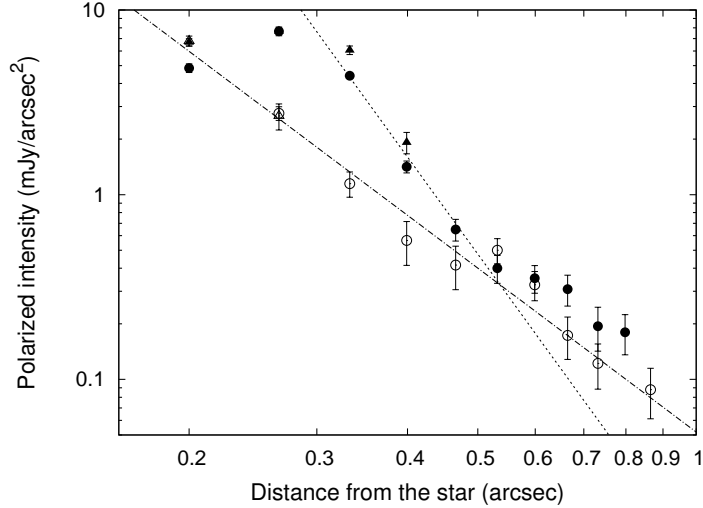} 
 \end{center}
\caption{Radial surface brightness distribution in $PI$. The SE (PA =$119\arcdeg$) profile is shown with $filled$ $circles$  (sPDI) and $triangles$ (qPDI). The NW (PA=$299\arcdeg$) profile is presented with $open$ $circles$ (sPDI) and $triangles$ (qPDI). The $dotted$ line for the SE region shows the relation of $r^{-5.4 \pm 0.5}$ estimated in $r=$$0\farcs30$--$0\farcs60$ whereas the {\it dashed-dotted} line is proportional to $r^{-3.0 \pm 0.5}$ measured in $r=0\farcs24$--$0\farcs55$. The errorbars show the uncertainties estimated using the images of $Q$ and $U$ ($\sigma$).}\label{fig:sb}
\end{figure}

Figure~\ref{fig:sb} shows the radial dependence of the surface brightness of the $PI$ in the SE (P.A. = $119\arcdeg$) and NW (P.A. = $299\arcdeg$) directions. 
The brightness was measured using the data binned over 7$\times$7 pixels (1 FWMH of the PSF). 
Overall, the distribution is in agreement between the sPDI and qPDI modes, and the peak flux density in SE is well matched. 
The least-squares fitting was performed using the sPDI data with a higher SNR than that of the qPDI, and  yielded the power-law relation of $r^{-3.0 \pm 0.5}$ for the NW direction in $0\farcs24 \leq r \leq 0\farcs55$. On the other hand, the $PI$ in the SE region including the arc was proportional to $r^{-5.4 \pm 0.5}$ in $0\farcs30 \leq r \leq 0\farcs60$. 
We confirmed that the uncertainty associated with the subtracted polarized PSF ($P_{\rm PSF} = 0.85 \pm 0.05$\%, $\theta_{\rm PSF} = 29\arcdeg \pm 2\arcdeg$) did not significantly alter the radial slope. 

The SE arc has its brightness peak at $r=0\farcs27$ at a P.A. of 119$\arcdeg$.  
Beyond this peak, the radial distribution shows a steeper decline in the SE than the $r^{-3}$ measured in the NW. The radial index of $-3$ is expected for a vertically flat disk or a disk with a constant opening angle \citep{whi92,mom15}. 
This slope has commonly been observed for protoplanetary disks in Stokes $I$, and has been found in $PI$ as well (e.g., \cite{tan12}). 
Therefore, we propose that the region behind the arc is vertically settled and shadowed by the arc structure. 
The same characteristic has been observed for the Herbig Ae star MWC~758; the distinct spiral arm was imaged in the southeastern region and the $PI$ dropped as $r^{-5.7}$ behind the radial brightness peak of the arm \citep{gra13}. 

\subsection{Arc-like structure}
The outer ($r$$\gtrsim$50~au) disk of V1247~Ori is characterized by an arc-like feature detected within $\sim$200~au of the star. 
The disks consisting of two arcs facing each other have been known (e.g., \cite{has12}), 
but in the case of V1247~Ori, no counterpart for the SE arc exists in NW at the same or similar distance from the star. 
Meanwhile, several Herbig Fe/Ae/Be stars, such as  SAO~206462 and MWC~758 \citep{mut12,gra13}, show  conspicuous arms at separations of $\lesssim$200~au in the shapes of spirals, where the distance from the star varies with azimuthal angle. 
Arm-like structures suggesting a deviation from an inclined circular disk have also been observed for Oph IRS 48 and HD~100546 \citep{fol15,ave14b,cur14}. 

\begin{figure}
 \begin{center}
  \includegraphics[width=80mm]{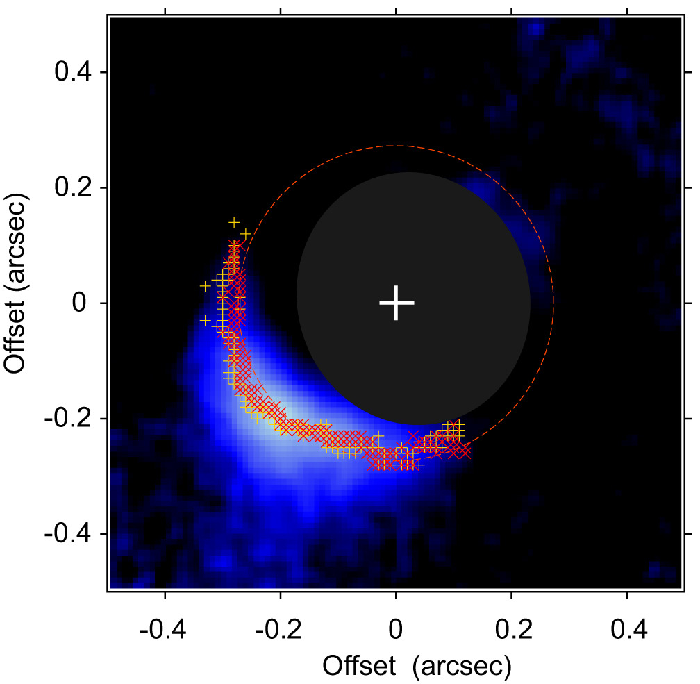}
  \hspace{5mm}
  \includegraphics[width=80mm]{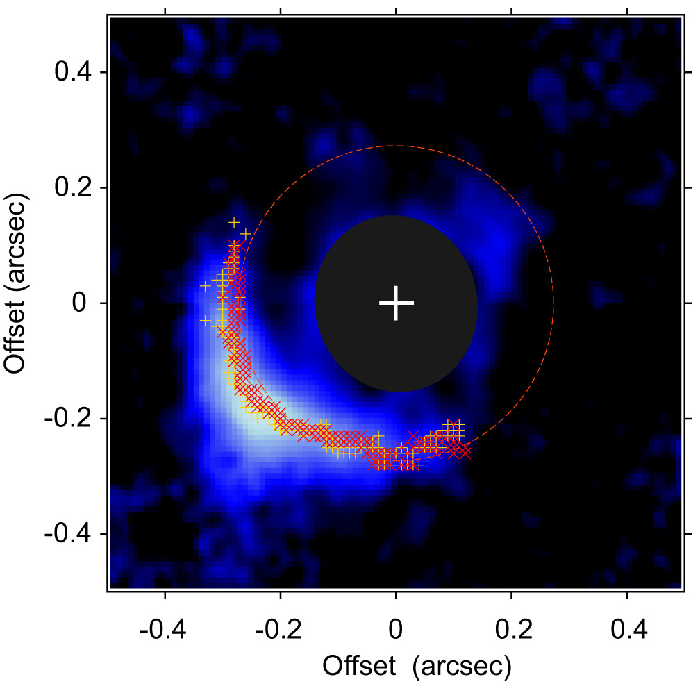}
 \end{center}
\caption{Arc structure traced by the radial maximum brightness, overlaid on the $PI$ image corrected for the disk inclination and the reduction of the incident stellar flux. The inclination and P.A. of the major axis are assumed to be $23\fdg8$ and $104\arcdeg$, respectively. The $left$ panel shows the image obtained in sPDI and the $right$ one is in qPDI. In each panel, the positions are marked with symbols both for sPDI ($red$  $cross$ ($\times$)) and qPDI ($yellow$ $plus$ (+)) for comparison, and the {\it dashed line} shows the fitted circle centered on the star. The stellar location is denoted with a cross.}\label{fig:circle-fit}
\end{figure}

We measured the shape of the arc for V1247~Ori as follows. 
The disk was first de-projected using the following inclinations and position angles: 
the P.A. of the major axis was $104\arcdeg$ or $119\arcdeg$, and the inclination was $31\fdg3$ or, at the end of the estimated range, $38\fdg8$ or $23\fdg8$ (\cite{kra13}, see also Section 4.1). 
Next, to compensate for the drop-off of the stellar flux incident upon the scatterer, the $PI$ at each location was multiplied by $r^2$ assuming no extinction during the travel of the incident photon. 
Then, the location of the maximum brightness in the radial profile was identified at each azimuthal angle in steps of $1\arcdeg$ to define the shape of the arc. 
The measurement was done in the azimuthal range where the $PI$ was detected over $3\sigma$ (Table~\ref{tab:circle_fit}). 
Figure~\ref{fig:circle-fit} shows the maximum spine determined by this method. 
The Gaussian fitting was also performed for the radial profile at each P.A. using $\sigma$ as a weight, although the profile was not well approximated as Gaussian in particular near the azimuthal edges of the arc.
The difference between the maximum brightness and the Gaussian-fitted peak was $0\farcs02$ on average. 
This $0\farcs02$ was regarded as the uncertainty in the arc position. 
The distance to the arc from the star was almost constant considering this positional uncertainty for all six cases, indicating consistency with a circular arc, rather than a spiral arm.
The shape of the arc determined using the sPDI images matches that obtained in qPDI within an uncertainty of $0\farcs02$. 

Least-squares fitting of a circle was carried out for the measured spine of the arc. 
The center of the circle was fixed at the stellar position in the first attempt, and the results are shown in Table~\ref{tab:circle_fit}. 
The obtained radii of the fitted circle agreed within the uncertainty for the six assumed cases of inclination angle and P.A. of the major axis. 
The average of these radii is $0\farcs28 \pm 0\farcs09$ (108 $\pm$ 35~au). 
The arc is  not necessarily centered on the star; thus, the fitting was also performed by varying the central position of the circle. 
Columns 6 and 7 denoted by $\Delta r$ and $\Delta$P.A. in Table~\ref{tab:circle_fit} show the deviations of the arc center from the stellar location, which were significant for all six cases. 
However, the direction of the shift depended on the assumed angles of the major axis and inclination. 
There was a trend toward the northern direction, but this direction was not always along the minor or major axes, and it is hard to relate this possible displacement to a disk geometry (a vertical structure; e.g., \cite{vin03}) or any mechanism for creating the arc (e.g., \cite{mit15}). 
Therefore, we cannot explicitly conclude an off-center for the arc, and its presence will need be investigated in the future once the disk geometry (major axis and inclination) is  accurately determined for the outer disk, for example through molecular-line observations. 

Note that we cannot exclude the possibility of a spiral arm with a small pitch angle that is hard to discriminate from a circular arc within the positional uncertainty; our data would be consistent with a spiral arm with a pitch angle of  $\lesssim3\arcdeg$. 
Another source of uncertainty is the disk vertical height which was ignored in the above fitting procedure. 
Although no measurements on the disk scale height are available, if assuming the disk aspect ratio ($H/r$ where $H$ is the height of the scattering surface of the disk) of 0.1 at $r=108$~au and the inclination of $31\fdg3$, then the arc is not a circle but a spiral arm with a pitch angle of about $2\fdg5$ (e.g., \cite{vin03}).

\begin{table}
  \caption{Circle fitting}\label{tab:circle_fit}
  \begin{center}
    \begin{tabular}{ccc||c|ccc}
      \hline
      \multicolumn{1}{c}{major axis}  & & & \multicolumn{1}{|c|}{centered on the star} & \multicolumn{3}{|c}{free center}   \\
      P.A. ($\arcdeg$) & $i$ ($\arcdeg$) & P.A. range ($\arcdeg$) & best-fit radius ($\arcsec$) & best-fit radius ($\arcsec$)  & $\Delta r$ ($\arcsec$) & $\Delta$P.A. ($\arcdeg$)\\ \hline
      104 & 23.8 & 67--206 & 0.273$\pm$0.019 & 0.269$\pm$0.004 & 0.021$\pm$0.004 & 68 \\
          & 31.3 & 68--205 & 0.281$\pm$0.015 & 0.281$\pm$0.004 & 0.009$\pm$0.004 & 87 \\
          & 38.8 & 64--206 & 0.292$\pm$0.015 & 0.299$\pm$0.005 & 0.010$\pm$0.004 & 148 \\ \hline
      119 & 23.8 & 65--209 & 0.269$\pm$0.019 & 0.282$\pm$0.005 & 0.023$\pm$0.004 & 27 \\
      	  & 31.3 & 66--209 & 0.277$\pm$0.016 & 0.310$\pm$0.005 & 0.038$\pm$0.005 & 159 \\
      	  & 38.8 & 64--209 & 0.287$\pm$0.018 & 0.335$\pm$0.005 & 0.060$\pm$0.005 & 148 \\ \hline
 \multicolumn{1}{@{}l}{\rlap{{\small 
 Columns 6 and 7 show the positional difference of the center of the fitted circle relative to the star.}}}
    \end{tabular}
  \end{center}
\end{table}

\section{Discussion}
\subsection{Comparison with other (pre-)transitional disks}
Arm-like or arc structures at least in scattered light, are relatively common for (pre-)transitional disks around Herbig Fe/Ae/Be stars \citep{qua13,clo14,reg14,bil14}.
Two or more arms have normally been detected for one disk and a disk with only one distinct arc has yet to be discovered (SAO~206462, MWC~758, AB~Aur, HD~100546, HD~142527; \cite{mut12,gra13,has11,boc13,cas12}). 
Oph IRS 48 may have an arm but two additional arcs were also detected in $PI$ \citep{fol15}. 
It is possible that V1247~Ori, which is a more distant object than those imaged so far, has other arms in the inner region not observable in our imaging ($r\lesssim$50~au); however, the azimuthal contrast in scattered light detected in our V1247~Ori imaging is higher than in any other disks imaged so far because of the one-arc structure. 
The SE part of the disk is 4 times brighter than the NW part in the integrated $PI$ (Section~4.1).
Comparing the peak brightness of the SE arc (P.A. = 199$\arcdeg$) with that of the NW disk (P.A. = 299$\arcdeg$) at the same distance from the star ($r=0\farcs27$), the arc was found to be 2.8 times brighter (Figure~\ref{fig:sb}). 
The brightness asymmetry along the major axis has been reported for the arcs of Oph~IRS~48 and HD~100546 just outside of the holes (gaps), but the contrast is weaker than V1247~Ori for both disks (Figure 5 in \cite{fol15}, Figure 4 in \cite{ave14b}). 

\begin{figure}
 \begin{center}
  \includegraphics[width=5.8cm,angle=-90]{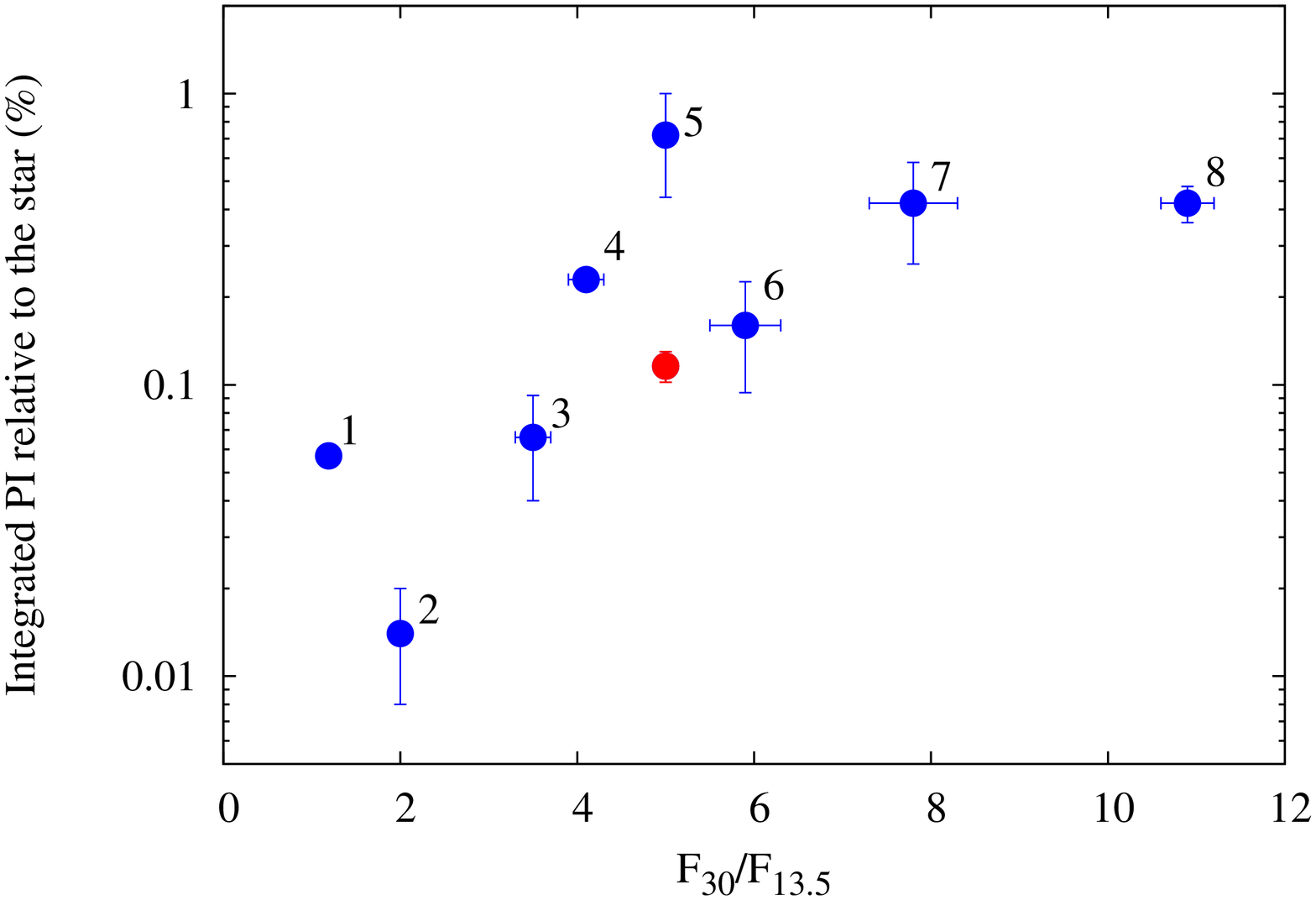} 
  \includegraphics[width=5.8cm,angle=-90]{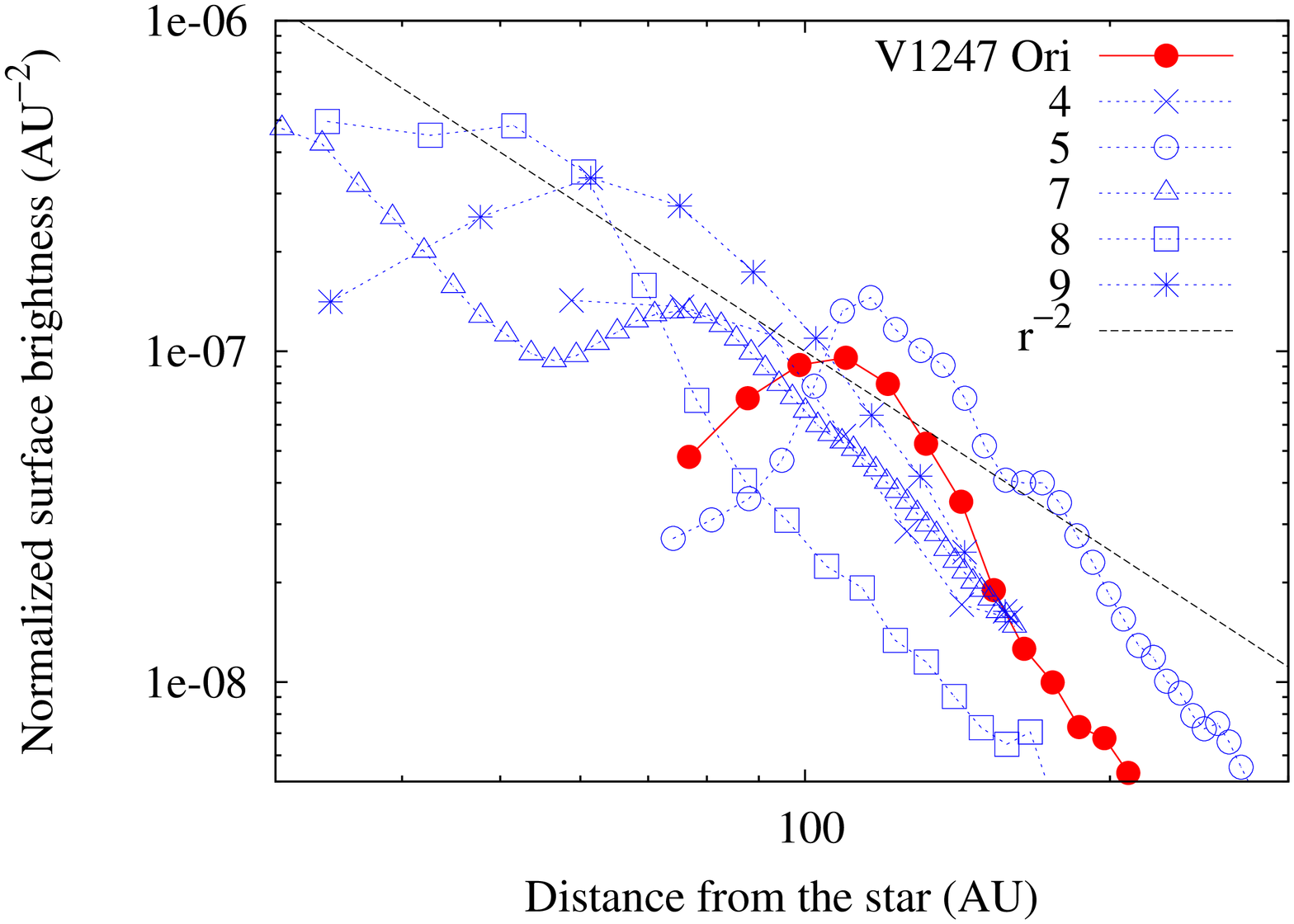} 
 \end{center}
\caption{Comparison with other disks in $PI$. The $PI$ is shown in $red$ for V1247~Ori, and in $blue$ for others. $Left$: Integrated polarized intensity normalized by the total ($\sim$stellar) brightness plotted against the flux ratio between 30 and 13.5~$\mu$m \citep{ack10,kra13}.  The sources in this panel are (1) MWC~480 \citep{kus12}, (2) HD~163296 \citep{gar14}, (3) HD~100546 \citep{qua11}, (4) MWC~758 \citep{gra13}, (5) HD~142527 \citep{ave14a}, (6) HD~97048 \citep{qua12}, (7) HD~169142 \citep{qua13,mom15}, and (8) SAO~206462 \citep{mut12}. Note that the observed radial range differs among these sources depending primarily on the distance. $Right$: Surface brightness normalized by the total ($\sim$stellar) flux density  expressed per square au. The dashed line shows a guide line for the proportionality of $\propto r^{-2}$. The P.A.s for the measurements are 90$\arcdeg$ for MWC~758 (4, \cite{gra13}), $-90\arcdeg$ for HD~142527 (5, \cite{ave14a}), $235\arcdeg$ for SAO~206462 (8, \cite{mut12}), and $97\fdg4$ for Oph IRS 48 (9, \cite{fol15}).  The profile is azimuthally averaged for HD~169142 (7, \cite{mom15}). The distance of MWC~758 is assumed to be 280~pc in this plot, whereas 200~pc has also been suggested \citep{gra13}.}\label{fig:comp_scat}
\end{figure}

To confirm whether the disk of V1247 Ori has another unique property, the level of $PI$ was compared with other disks surrounding stars of $\sim$$2M_{\solar}$. 
The left panel of Figure~\ref{fig:comp_scat} shows the $PI$ integrated over the detected area plotted against the flux ratio between 30 and 13.5~$\mu$m ($F_{30}/F_{13.5}$). 
The latter quantity has been used for the SED classification originally proposed by \citet{mee01}, where sources fall into either group I or II. 
This classification is sensitive to the presence and size of an inner hole/gap at the corresponding temperature of a few to several hundred K, and the degree of disk vertical flaring \citep{ack10,maa13}. 
The two objects with the smallest $F_{30}/F_{13.5}$ are in group II whereas others are in group I and are known to be (pre-)transitional disks with large (tens of au) gaps \citep{kus12,mut12,gra13,qua13,gar13,gar14,mom15}. 
Note that the group II disks are variable in scattered light and that MWC~480 at $F_{30}/F_{13.5}=1.19$ is considered to be in a bright phase in the measurements \citep{kus12}. 
The comparison of the $PI$s is not quite straightforward, because the disks have a range of inclinations (10$\arcdeg$--60$\arcdeg$) and spatial structures, resulting in different scattering angles. 
In addition, V1247~Ori is more distant than the others (100--140~pc), and the integrated $PI$ should be considered as a lower limit at  the comparison. 
An increase by more than one order of magnitude is unlikely, however, when considering the $PI$ measured from $\sim$30~au ($0\farcs2$ at 140~pc) to $\sim$60~au for other group I disks. 
Thus, V1247~Ori shows a reasonable $PI$, that does not disrupt the established tendency for larger $F_{30}/F_{13.5}$ to correspond to brighter scattered light from the disk, as was also reported by \citet{gar14}. 

The right panel of Figure~\ref{fig:comp_scat} presents the radial surface brightness for the (pre-)transitional disks for which spirals or arc-like structures have been resolved in scattered light. The surface brightness is normalized by the total ($\sim$stellar) intensity and expressed per squared au. 
The profile was measured along the major axis for one disk with an inclination of $50\arcdeg$. 
The major-axis profiles are not always available in the literature for other disks, but because their inclinations are less than $\sim$30$\arcdeg$,  the effect of the inclination would likely be negligible compared to the contributions from the non-uniform disk structures. 
The $PI$ distribution peaks at the location of the arms (i.e., the inner rim of the outer disk), and those brightness peaks roughly follow a proportionality  of $\propto r^{-2}$.
This is plausible considering the inverse square relation of light attenuation, assuming that the incident light does not experience considerable extinction by the disk material inside the radial peak and is scattered-off at the surface of the optically-thick outer disk.
V1247~Ori is not an exception, and its peak brightness at the SE arc is consistent with this tendency. 
Therefore, the object is not an outlier in terms of the absolute flux density in $PI$, and the grains in its disk seem to have similar scattering properties as those observed in other (pre-)transitional disks.

\subsection{Spatial structure of the disk}
Why do we see an arc only in the SE? 
The outer disk of V1247~Ori is likely to be optically thick given the abundant infrared excess, and hence the scattered light imaging in the near-infrared is only sensitive to the structure at the disk surface. 
The brightness in scattered light reflects how well the disk is illuminated by the star (grazing angle); the inner rim of the SE arc should be more readily exposed by the stellar light than the NW region, as discussed in Section 5.1. 

\citet{kra13} found that the gap between the inner disk ($r=0.18$--0.27~au) and the outer disk ($\gtrsim$46~au) was not completely empty and that the distribution of material within the gap was not spatially uniform. 
The optical thickness for the gap material is not tightly constrained and it could partly be optically thick \citep{kra13}. 
If an optically thick structure exists in the NW, it could act as a wall blocking the stellar light and casting a shadow onto the outer region. 
Alternatively, the shadowing might be because of the inner wall of the outer disk at around 46~au if the wall is vertically more flared to the NW than to the SE. 
In these cases, it is possible that the disk at about 108~au is in a ring-like shape, or that the outer disk continuously extends from 46 to 108 au as a density structure, but that only the SE part is detected in scattered light.
A drawback of this hypothesis is that the radial dependence of $PI$ observed in NW, $\propto r^{-3}$, cannot easily be reconciled with the shadowing, which predicts a steeper decline. 

A non-axisymmetric density structure 
in the outer disk could also cause the SE--NW contrast. 
Our imaging cannot exclude the possibility that, in terms of density (spatial structure), the outer disk of V1247 Ori consists of an arc on one side, in SE, and a background disk without distinct structure. 
The question is what physical mechanism could be responsible for this semi-circular density distribution. 
Recent sub- and millimeter observations have revealed the density enhancement of grains with large opacity at those wavelengths ($\sim$millimeter-sized) in both the radial \citep{wal14} and azimuthal directions \citep{cas12,fuk13,mar13}. 
Such a disk with strong (over a factor $\sim$10) azimuthal asymmetry looks like a crescent, which may be approximated by an arc. 
The ongoing discussion suggests that the structure can be explained, for instance, by dust trapping in gas vortices \citep{lyr09,bir13} or gas drag in an arm developed in the mode of azimuthal wavenumber $m=1$ \citep{mit15}. 
However, based on the two known cases (Oph IRS 48 and HD~142527), their asymmetry is less prominent in scattered light and scattering in polarized light was also detected at azimuthal angles  not associated with the submillimeter crescents \citep{ave14a,fol15}. 
This is probably because a wall-like structure, which is the outer boundary of the hole or gap, exists in the gaseous component over nearly the whole range of azimuthal angles \citep{bru14,mut15} and small grains mixed with gas are detectable in scattered light. 
This is not the situation that we found for V1247~Ori. 
This difference can be interpreted as a gaseous wall at 108~au, which is much less conspicuous on the NW side than on the SE and compared to the other disks.

An alternate explanation is that the arc is a spiral arm caused by several possible mechanisms, including the presence of a companion with planetary mass as well as the marginally unstable, self-gravitating disk. 
Given that the spirals for other sources also consist of nearly circular regions in the observed scattered-light images \citep{mut12,gra13,ave14b,ben15}, we cannot exclude the possibility that only the circular part was detected in our imaging and that the tail/head was undetected because of the low sensitivity in $PI$ in the direction of the disk's minor axis. 
Note that, regardless of whether the arc is a spiral arm or not, or whether it is driven by a companion or not, it may be reasonable to expect dust over-density in the disk mid-plane just behind the illuminated inner wall of the arc seen in scattered light.

In conclusion, the asymmetric structure suggests the presence of unseen companions and/or local density-enhancement around the mid-plane, which may provide the ideal site for planet formation. 
In searching for companion objects, the parameter space for the mass ratio $\lesssim0.1$ is still yet to be explored \citep{kra13}, and high-contrast observations will be important for making conclusions about the physical nature of the arc. 
Radio interferometric observations to determine the outer disk geometry will be useful to more accurately determine the arc shape and to identify any displacement of the center of the SE arc from the star. 
More importantly, high-angular resolution observations with ALMA are essential for understanding the disk nature by obtaining the density distribution both in gas and in dust at the mid-plane, as well as the temperature structure within the disk.

If there is a ring at 108~au, a radial gap could exist between this ring and the disk component with an inner boundary at $\sim$46~au, where it was estimated to exist on the basis of mid-infrared interferometric observations \citep{kra13}. 
A planetary companion is one plausible explanation for a radial gap in general. 
Although significant uncertainty is expected in the width of the gap, an upper limit of the gap width can be provided by the lowest-order (2:1) Lindblad resonance (e.g., \cite{tak96}). 
If this is the case, and assuming the outer gap radius of 108~au, one can infer that a putative companion is located at 68~au and that the inner radius of the gap is at 43~au, which is close to the observed inner edge of the outer disk at 46~au. 
This discussion is quite speculative at present, considering the uncertainty in the position measurements (108 $\pm$ 35~au), but it would be worth mentioning that the location of the detected arc can be related to the inner disk component previously found \citep{kra13} in this context.

\section{Summary}
The circumstellar disk of V1247~Ori was spatially resolved in the $H$-band using polarimetric differential imaging at a resolution of $0\farcs07$ (27~au). 
The region beyond $0\farcs14$ (54~au) was observed and the polarized light was detected out to a distance of $0\farcs86$ (330~au) in the SE (P.A. = 119$\arcdeg$) and $0\farcs66$ (254~au) in the NW (P.A. = 299$\arcdeg$). 
A remarkable arc structure was discovered at the P.A. range from about $60\arcdeg$ to $210\arcdeg$. 
The radial surface brightness distribution was consequently found to be asymmetric, with a $PI$  proportional to $r^{-5.4}$ to the SE behind the peak of the arc ($r \geq 0\farcs30$), while showing a monotonic decrease as $r^{-3.0}$ to the NW. 
The one, conspicuous arc is responsible for the azimuthal contrast, which is exceptionally large compared to any other (pre-)transitional disks imaged in scattered light so far.
The shape of the arc is consistent with an arc at a distance of $0\farcs28 \pm 0\farcs09$ (108 $\pm$ 35~au) from the  star, although the possibility of a spiral with a small pitch angle ($\lesssim3\arcdeg$) cannot be excluded. 
Although other possible explanations exist, unseen companions could be the reason for the observed asymmetry, 
and hence V1247~Ori is another ideal object for investigating the formation mechanism of companion bodies with mass ratios of $\lesssim0.1$, as well as the mutual evolution of a protoplanetary disk and putative perturbers.

\bigskip

This work is supported by MEXT KAKENHI Nos. 23103005. S.K. acknowledges support from an STFC Ernest Rutherford Fellowship (ST/J004030/1) and Marie Curie CIG grant (SH-06192).



\begin{thebibliography}{}
\bibitem[Acke et al.(2010)]{ack10} Acke, B., Bouwman, J., Juh{\'a}sz, A., et al.\ 2010, \apj, 718, 558 
\bibitem[Andrews et al.(2011)]{and11} Andrews, S.~M., Wilner, D.~J., Espaillat, C., et al.\ 2011, \apj, 732, 42 
\bibitem[Avenhaus et al.(2014a)]{ave14a} Avenhaus, H., Quanz, S.~P., Schmid, H.~M., et al.\ 2014, \apj, 781, 87  
\bibitem[Avenhaus et al.(2014b)]{ave14b} Avenhaus, H., Quanz, S.~P., Meyer, M.~R., et al.\ 2014, \apj, 790, 56 
\bibitem[Benisty et al.(2015)]{ben15} Benisty, M., Juhasz, A., Boccaletti, A., et al.\ 2015, \aap, accepted (arXiv:1505.05325)
\bibitem[Biller et al.(2014)]{bil14} Biller, B.~A., Males, J., Rodigas, T., et al.\ 2014, \apjl, 792, LL22 
\bibitem[Birnstiel et al.(2013)]{bir13} Birnstiel, T., Dullemond, C.~P., \& Pinilla, P.\ 2013, \aap, 550, LL8 
\bibitem[Boccaletti et al.(2013)]{boc13} Boccaletti, A., Pantin, E., Lagrange, A.-M., et al.\ 2013, \aap, 560, AA20 
\bibitem[Brown et al.(2009)]{bro09} Brown, J.~M., Blake, G.~A., Qi, C., et al.\ 2009, \apj, 704, 496 
\bibitem[Bruderer et al.(2014)]{bru14} Bruderer, S., van der Marel, N., van Dishoeck, E.~F., \& van Kempen, T.~A.\ 2014, \aap, 562, AA26 
\bibitem[Caballero(2010)]{cab10} Caballero, J.~A.\ 2010, \aap, 511, LL9 
\bibitem[Caballero(2008)]{cab08} Caballero, J.~A.\ 2008, \mnras, 383, 750 
\bibitem[Calvet et al.(2002)]{cal02} Calvet, N., D'Alessio, P., Hartmann, L., et al.\ 2002, \apj, 568, 1008 
\bibitem[Casassus et al.(2012)]{cas12} Casassus, S., Perez M., S., Jord{\'a}n, A., et al.\ 2012, \apjl, 754, LL31 
\bibitem[Casassus et al.(2013)]{cas13} Casassus, S., van der Plas, G., M, S.~P., et al.\ 2013, \nat, 493, 191 
\bibitem[Currie et al.(2014)]{cur14} Currie, T., Muto, T., Kudo, T., et al.\ 2014, \apjl, 796, LL30 
\bibitem[Casassus et al.(2012)]{cas12} Casassus, S., Perez M., S., Jord{\'a}n, A., et al.\ 2012, \apjl, 754, LL31 
\bibitem[Close et al.(2014)]{clo14} Close, L.~M., Follette, K.~B., Males, J.~R., et al.\ 2014, \apjl, 781, LL30 
\bibitem[Cutri et al.(2003)]{cut03} Cutri, R.~M., Skrutskie, M.~F., van Dyk, S., et al.\ 2003, VizieR Online Data Catalog, 2246, 0 
\bibitem[Follette et al.(2015)]{fol15} Follette, K.~B., Grady, C.~A., Swearingen, J.~R., et al.\ 2015, \apj, 798, 132 
\bibitem[Fukagawa et al.(2013)]{fuk13} Fukagawa, M., Tsukagoshi, T., Momose, M., et al.\ 2013, \pasj, 65, L14 
\bibitem[Fukagawa et al.(2010)]{fuk10} Fukagawa, M., Tamura, M., Itoh, Y., et al.\ 2010, \pasj, 62, 347 
\bibitem[Garufi et al.(2013)]{gar13} Garufi, A., Quanz, S.~P., Avenhaus, H., et al.\ 2013, \aap, 560, AA105 
\bibitem[Garufi et al.(2014)]{gar14} Garufi, A., Quanz, S.~P., Schmid, H.~M., et al.\ 2014, \aap, 568, AA40 
\bibitem[Grady et al.(2013)]{gra13} Grady, C.~A., Muto, T., Hashimoto, J., et al.\ 2013, \apj, 762, 48 
\bibitem[Hashimoto et al.(2011)]{has11} Hashimoto, J., Tamura, M., Muto, T., et al.\ 2011, \apjl, 729, LL17 
\bibitem[Hashimoto et al.(2012)]{has12} Hashimoto, J., Dong, R., Kudo, T., et al.\ 2012, \apjl, 758, LL19 
\bibitem[Hayano et al.(2004)]{hay04} Hayano, Y., Saito, Y., Saito, N., et al.\ 2004, \procspie, 5490, 1088 
\bibitem[Honda et al.(2012)]{hon12} Honda, M., Maaskant, K., Okamoto, Y.~K., et al.\ 2012, \apj, 752, 143 
\bibitem[Isella et al.(2013)]{ise13} Isella, A., P{\'e}rez, L.~M., Carpenter, J.~M., et al.\ 2013, \apj, 775, 30 
\bibitem[Joos et al.(2008)]{joo08} Joos, F., Buenzli, E., Schmid, H.~M., \& Thalmann, C.\ 2008, \procspie, 7016, 70161I 
\bibitem[Yamamoto et al.(2013)]{yam13} Yamamoto, K., Matsuo, T., Shibai, H., et al.\ 2013, \pasj, 65, 90 
\bibitem[Kraus et al.(2013)]{kra13} Kraus, S., Ireland, M.~J., Sitko, M.~L., et al.\ 2013, \apj, 768, 80 
\bibitem[Kusakabe et al.(2012)]{kus12} Kusakabe, N., Grady, C.~A., Sitko, M.~L., et al.\ 2012, \apj, 753, 153 
\bibitem[Lyra et al.(2009)]{lyr09} Lyra, W., Johansen, A., Klahr, H., \& Piskunov, N.\ 2009, \aap, 493, 1125 
\bibitem[Maaskant et al.(2013)]{maa13} Maaskant, K.~M., Honda, M., Waters, L.~B.~F.~M., et al.\ 2013, \aap, 555, A64 
\bibitem[Mayama et al.(2012)]{may12} Mayama, S., Hashimoto, J., Muto, T., et al.\ 2012, \apjl, 760, L26 
\bibitem[Meeus et al.(2001)]{mee01} Meeus, G., Waters, L.~B.~F.~M., Bouwman, J., et al.\ 2001, \aap, 365, 476 
\bibitem[Mittal \& Chiang(2015)]{mit15} Mittal, T., \& Chiang, E.\ 2015, \apjl, 798, LL25 
\bibitem[Momose et al.(2015)]{mom15} Momose, M., Morita, A., 
Fukagawa, M., et al.\ 2015, \pasj, 67, 83 
\bibitem[Muto et al.(2012)]{mut12} Muto, T., Grady, C.~A., Hashimoto, J., et al.\ 2012, \apjl, 748, LL22 
\bibitem[Muto et al.(2015)]{mut15} Muto, T., Tsukagoshi, T., Momose, M., et al.\ 2015, \pasj, 67, 122 
\bibitem[Quanz et al.(2011)]{qua11} Quanz, S.~P., Schmid, H.~M., Geissler, K., et al.\ 2011, \apj, 738, 23 
\bibitem[Quanz et al.(2012)]{qua12} Quanz, S.~P., Birkmann, S.~M., Apai, D., Wolf, S., \& Henning, T.\ 2012, \aap, 538, A92 
\bibitem[Quanz et al.(2013)]{qua13} Quanz, S.~P., Avenhaus, H., Buenzli, E., et al.\ 2013, \apjl, 766, LL2 
\bibitem[Reggiani et al.(2014)]{reg14} Reggiani, M., Quanz, S.~P., Meyer, M.~R., et al.\ 2014, \apjl, 792, LL23 
\bibitem[Rodigas et al.(2014)]{rod14} Rodigas, T.~J., Follette, K.~B., Weinberger, A., Close, L., \& Hines, D.~C.\ 2014, \apjl, 791, L37 
\bibitem[Sitko et al.(2012)]{sit12} Sitko, M.~L., Day, A.~N., Kimes, R.~L., et al.\ 2012, \apj, 745, 29 
\bibitem[Strom et al.(1989)]{str89} Strom, K.~M., Strom, S.~E., Edwards, S., Cabrit, S., \& Skrutskie, M.~F.\ 1989, \aj, 97, 1451 
\bibitem[Suzuki et al.(2010)]{suz10} Suzuki, R., Kudo, T., Hashimoto, J., et al.\ 2010, \procspie, 7735, 773530 
\bibitem[Takami et al.(2014)]{tak14} Takami, M., Hasegawa, Y., Muto, T., et al.\ 2014, \apj, 795, 71 
\bibitem[Takeuchi et al.(1996)]{tak96} Takeuchi, T., Miyama, S.~M., \& Lin, D.~N.~C.\ 1996, \apj, 460, 832 
\bibitem[Tamura(2009)]{tam09} Tamura, M.\ 2009, American Institute of Physics Conference Series, 1158, 11 
\bibitem[Tamura et al.(2006)]{tam06} Tamura, M., Hodapp, K., Takami, H., et al.\ 2006, \procspie, 6269, 62690V 
\bibitem[Tanii et al.(2012)]{tan12} Tanii, R., Itoh, Y., Kudo, T., et al.\ 2012, \pasj, 64, 124 
\bibitem[Terrell et al.(2007)]{ter07} Terrell, D., Munari, U., \& Siviero, A.\ 2007, \mnras, 374, 530 
\bibitem[van der Marel et al.(2002)]{van02} van der Marel, R.~P., Gerssen, J., Guhathakurta, P., Peterson, R.~C., \& Gebhardt, K.\ 2002, \aj, 124, 3255 
\bibitem[van der Marel et al.(2013)]{mar13} van der Marel, N., van Dishoeck, E.~F., Bruderer, S., et al.\ 2013, Science, 340, 1199 
\bibitem[Vieira et al.(2003)]{vie03} Vieira, S.~L.~A., Corradi, W.~J.~B., Alencar, S.~H.~P., et al.\ 2003, \aj, 126, 2971 
\bibitem[Vinkovi{\'c} et al.(2003)]{vin03} Vinkovi{\'c}, D., Ivezi{\'c}, {\v Z}., Miroshnichenko, A.~S., \& Elitzur, M.\ 2003, \mnras, 346, 1151 
\bibitem[Walsh et al.(2014)]{wal14} Walsh, C., Juh{\'a}sz, A., Pinilla, P., et al.\ 2014, \apjl, 791, LL6 
\bibitem[Whitney \& Hartmann(1992)]{whi92} Whitney, B.~A., \& Hartmann, L.\ 1992, \apj, 395, 529 

\end{thebibliography}
\end{document}